\documentstyle[12pt]{article}

\newcommand{\ann}{\rm ann}		% annihilation
\newcommand{\bt}{\rm b}			% bottom quark
\newcommand{\crit}{\rm c}		% critical
\newcommand{\D}{\rm D}			% decoupled
\newcommand{\dn}{\rm d}			% down quark; differentiation sign
\newcommand{\Dn}{\rm D}			% Down quark ... generic
\newcommand{\e}{\rm e}			% electron
\newcommand{\E}{\rm E}			% Electron ... generic
\newcommand{\eq}{\rm eq}		% equilibrium
\newcommand{\f}{\rm f}			% fermion
\newcommand{\fr}{\rm fr}		% freeze-out
\newcommand{\GeV}{\rm GeV}		% GeV
\newcommand{\GUT}{\rm GUT}		% GUT
\newcommand{\h}{\rm h}			% Higgs boson
\newcommand{\I}{\rm I}			% interacting
\newcommand{\Lf}{\rm L}			% left
\newcommand{\MS}{\overline{\rm MS}}	% Minimal Subtraction
\newcommand{\Pl}{\rm P}			% Planck
\newcommand{\Q}{\rm Q}			% charge
\newcommand{\R}{\rm R}			% right
\newcommand{\s}{\rm s}			% strong
\newcommand{\tp}{\rm t}			% top quark
\newcommand{\up}{\rm u}			% up quark
\newcommand{\Up}{\rm U}			% Up quark ... generic
\newcommand{\X}{\rm X}			% X boson
\newcommand{\W}{\rm W}			% W boson
\newcommand{\Y}{\rm Y}			% hypercharge
\newcommand{\Z}{\rm Z}			% Z boson
\newcommand{\leqsim}{\,\raisebox{-0.6ex}{$\buildrel < \over \sim$}\,}
\newcommand{\geqsim}{\,\raisebox{-0.6ex}{$\buildrel > \over \sim$}\,}

\begin{document}

\pagestyle{empty}
\begin{flushright}
 OUTP-92-10P \\
 September 1992
\end{flushright}
\vspace{5mm}
\begin{center}
{\LARGE Neutralino Dark Matter \\ \vspace{5mm} in a Class of Unified Theories}
 \\ \vspace{1.5cm}
{\large S.A.~Abel and S.~Sarkar} \\ \vspace{5mm}
 Theoretical Physics, \\ University of Oxford, \\ Oxford OX1 3NP, U.K. \\
 \vspace{1cm}
{\large I.B.~Whittingham} \\ \vspace{5mm}
 Department of Physics \\ James Cook University \\ Townsville, Australia 4811
\\
\end{center}
\vspace{1cm}
\begin{abstract}
\noindent
The cosmological significance of the neutralino sector is studied for a
class of supersymmetric grand unified theories in which electroweak
symmetry breaking is seeded by a gauge singlet. Extensive use is made
of the renormalization group equations to significantly reduce the
parameter space, by deriving analytic expressions for all the
supersymmetry-breaking couplings in terms of the universal gaugino mass
$m_{1/2}$, the universal scalar mass $m_{0}$ and the coupling $A$. The
composition of the lightest supersymmetric partner is determined
exactly below the $\W$ mass, no approximations are made for sfermion
masses, and all particle exchanges are considered in calculating the
annihilation cross-section; the relic abundance is then obtained by an
analytic approximation. We find that in these models, stable
neutralinos may make a significant contribution to the dark matter in
the universe.
\end{abstract}
\vspace{1.5cm}
\begin{center}
(to appear in {\sl Nuclear Physics B})
\end{center}

\newpage
\pagestyle{plain}

\section{ Introduction }

The neutralino ($\chi$) in supersymmetric theories is a leading
candidate for the constituent of the dark matter in the universe
\cite{ellis1}. It has been demonstrated to be the lightest
supersymmetric partner (LSP), and to have a significant cosmological
relic density, in both the minimal supersymmetric standard model (MSSM)
\cite{goldberg}-\cite{drees}, and in an extended version \cite{ellis3}
in which symmetry breaking is driven by a gauge singlet
\cite{greene}-\cite{olive2}. In contrast to other dark matter
candidates such as the axion or gravitino, the neutralino has a relic
density $\rho_{\chi}$ such that its contribution to the cosmological
density parameter, $\Omega_{\chi} \equiv \rho_{\chi}/\rho_{\crit}$, is
naturally close to unity \cite{ellis1}. Here $\rho_{\crit} \simeq 1.05
\times 10^{-5} h^{2}$ GeV cm$^{-3}$ is the critical density
corresponding to a flat universe and $h \equiv H_{0} / 100$ km s$^{-1}$
Mpc$^{-1}$ is the (conventionally scaled) Hubble parameter, which is
observationally constrained to lie in the range $0.4 \leqsim h \leqsim
1.0$ \cite{kolb}.
\smallskip \par
The actual value of $\Omega$ is rather uncertain. Luminous matter, e.g.
in the disks of galaxies, contributes at most $\Omega \sim 0.01$ while
studies of dark matter in galactic halos and in groups and clusters of
galaxies imply $\Omega \sim 0.01 - 0.2$ \cite{binney}. Indirect
estimates of $\Omega$ based on observations of peculiar (i.e.
non-Hubble) motions on supercluster and larger scales suggest higher
values in the range $\sim 0.2 - 1$ \cite{kaiser}. The recent observations
by the COBE satellite of angular fluctuations in the microwave
background radiation has given strong support to the hypothesis that
the universe is dominated by dark matter with a density close to the
critical value; interpretations of these data require that for cold
dark matter, $\Omega \sim (0.2 - 0.5)\ h^{-1}$ \cite{wright}.
Additionally, the product $\Omega h^{2}\ (\propto \rho)$ can be
conservatively bounded from above by requiring that the age of the
universe exceed the observational lower limit of $\sim 10^{10}$ yr;
this yields $\Omega h^{2} \leqsim 1$ for $h \geqsim 0.4$, assuming a
Friedmann-Robertson-Walker cosmology \cite{kolb}. Further, if one
assumes that the dark matter provides the critical density, then the
age constraint restricts the Hubble parameter to be $h \leqsim 2/3$,
hence $ \Omega h^{2} \leqsim 4/9$.  Taking all this into account, we
adopt as the range of cosmological interest:

\begin{equation}
 0.1 \leqsim \Omega_{\chi}\,h^{2} \leqsim 0.5\ .
\end{equation}

\noindent
Note that the range $0.01 \leqsim \Omega_{\chi}\,h^{2} \leqsim 0.1$
would be relevant only to the dark matter in galactic halos. Obviously
the constituent of the cosmological (cold) dark matter may also be the
constituent of the halo dark matter (but not neccessarily vice
versa).
\smallskip \par
In the standard picture of the evolution of the universe \cite{kolb},
it is assumed that expansion is adiabiatic following an early period of
inflation. Massive particles `freeze out' of chemical equilibrium with
the thermal radiation-dominated plasma when their rate of annihilation
falls behind the rate of expansion; hence their relic abundance is {\em
inversely proportional} to the thermally-averaged annihilation
cross-section $\langle \sigma_{\ann}\,v \rangle$. In order to expedite
the calculation of this quantity in supersymmetric theories, most
authors resort to assumptions about the values of various parameters in
order to reduce the multi-dimensional parameter space (e.g. 20-dim in
the case of the MSSM !). One common assumption, motivated by grand
unified theories (GUT), is that the gaugino masses are related by their
renormalisation from a common value at an energy scale of ${\cal O}
(10^{16})$ GeV, viz.

\begin{equation}
\label{GUTvalue}
 m_{\Y} = \frac{5}{3} \left(\frac{g_{\Y}^{2}}{g_{2}^{2}}\right) m_{2}\ .
\end{equation}

\noindent
One can go further and note that {\em all} the low energy
couplings are then related by the renormalization group equations
(RGEs) and a few model-dependent parameters. Use of this fact is
however seldom made in the literature, because the RGEs are not soluble
analytically, and also because the large value suggested by experiment
for the mass of the top quark requires its Yukawa coupling to be
quite close to the unitarity limit in the MSSM, thus prohibiting any
acceptable approximations.
\smallskip \par
On the other hand, attention has recently been drawn
\cite{lopez,ellis2,drees} to the importance of using accurate
expressions for the low energy couplings when determining the relic
abundance. The model considered in these papers was the MSSM as derived
from supergravity-inspired GUTs, which has only 6 arbitrary parameters
defined at the unification scale. These are the universal scalar
($m_{0}$) and gaugino ($m_{1/2}$) masses, the Higgs coupling ($\mu$)
and the hidden sector parameters ($A$ and $B$) and the top quark mass
($m_{\tp}$). In principle, the remaining parameters such as the Higgs
vacuum expectation values (VEVs) are determined by the electroweak
breaking constraints, although in practice matters are complicated by
large radiative corrections to the effective potential from top and
stop diagrams \cite{ellis4}.
\smallskip \par
In this paper we shall carry out a similar analysis for the `minimally'
extended supersymmetric standard model. Running the RGEs analytically,
we have obtained polynomial approximations for all the parameters at
the weak scale in terms of their values at the GUT scale.  Comparison
with numerical evaluations show our solutions to be accurate to better
than $5\%$, even close to the unitarity limit. This reduces the
dimensionality  of the parameter space to 6: $m_{1/2}$, $m_{0}$, $A$,
the top quark Yukawa coupling $\lambda_{2_{\tp}}$, and two Yukawa
couplings involving the Higgs fields and gauge singlets, $\lambda_{7}$
and $\lambda_{8}$. Again the Higgs VEVs are determinable from the
electroweak symmetry breaking requirements. In the following section we
describe the model and introduce the dark matter candidate. Then we
review the prescription for analytically obtaining the relic density to
an accuracy better than $5\%$ and, using our RGE approximations,
calculate its value for various choices of the Higgs VEVs. We find that
the neutralino has a relic abundance $\Omega_{\chi} \sim {\cal O} (1)$
over most of the experimentally allowed parameter space.

\section{ The Minimally Extended Standard Model }

The model we shall consider is the MSSM extended by including a gauge
singlet to give the necessary Higgs mixing and thus break SU(2)$_{\Lf}$
$\otimes$ U(1)$_{\Y}$ \cite{derendinger,barr}. The relevant
superpotential is

\begin{eqnarray}
\label{lowsuperpot}
 {\cal W} &= &\lambda_{1} {\Dn}_{\R} ({\Dn}_{\Lf}^{\dagger} {\h}^{0} -
  {\Up}_{\Lf}^{\dagger}{\h}^{-}) + \lambda_{2} {\Up}_{\R}
({\Up}_{\Lf}^{\dagger}
   {\bar {\h}}^{0} - {\Dn}_{\Lf}^{\dagger} {\bar {\h}}^{+}) \\
 & & + \lambda_{3} {\E}_{\R} ({\E}_{\Lf}^{\dagger} {\h}^{0}
 - \nu_{\Lf}^{\dagger}{\h}^{-})
 + \lambda_{7} ({\h}^{0} {\bar {\h}}^{0} - {\h}^{-} {\bar {\h}}^{+}) \phi_{0}
 + \lambda_{8}\phi^{3}_{0}.
\end{eqnarray}

\noindent
The first three terms give masses to the (s)quarks and (s)leptons, such
that $\lambda_{1} \langle {\h} \rangle$, $\lambda_{2} \langle {\bar {\h}}
\rangle$ and $\lambda_{3} \langle {\h} \rangle$ are the mass matrices of
the down quarks, up quarks, and electrons respectively (the generation
indices have been supressed). The form of eq.~(\ref{lowsuperpot}) is
clearly similar to the Standard Model superpotential, apart from the
presence of the last two couplings involving the gauge singlet
$\phi_{0}$, which replace the usual quadratic Higgs coupling, and
ensure a satisfactory breaking of SU(2)$_{\Lf}$ $\otimes$ U(1)$_{\Y}$. This
model is the minimal extension to the MSSM that contains only
trilinear couplings, thus making it a natural candidate to emerge from
level-one fermionic string theory. (It should be pointed out that in
the `flipped string' models \cite{anton1} no gauge singlet appears in
the low-energy effective potential, because such singlets may aquire
large VEVs at a scale of ${\cal O} (10^{11})$ GeV corresponding to the
condensation of scalars in the hidden SU(4) $\otimes$ SO(10) sector. We
are therefore not addressing schemes of this form.)
\smallskip \par
The phenomenology of such models has been studied extensively in the
literature \cite{greene}-\cite{olive2}, \cite{barr}-\cite{ellis7}, but
relatively little attention has been paid to the cosmological relic
abundance of the neutralinos therein. Other proposed dark matter
candidates are the `flatino' superpartner of the massless scalar flat
field (`flaton') after SU(5) $\otimes$ U(1) symmetry breaking
\cite{ellis6} and the integer-charged bound states (`cryptons') in some
hidden sector \cite{ellis7}. The flatino/neutralino mass matrix is
restricted by the amount of R-parity breaking which is induced. This is
because models with broken R-parity (such as the minimal flipped SU(5)
model considered in ref.~\cite{ellis6}) have relic particles which are
unstable. The decays of such particles into photons, charged particles
or even neutrinos may violate the astrophysical bounds obtained by
analysis of the diffuse $\gamma$-ray background and data from
underground and cosmic ray detectors \cite{ellis8}. Careful analysis of
the diagonalisation of the neutral fermion mass matrix shows
\cite{abel} that the mixing of the neutralino with the neutrino is
${\cal O} (m_{\W} / m_{\GUT})^{2}$, {\em not} the generic value of
${\cal O} (m_{\W} / m_{\GUT})$ adopted earlier \cite{ellis6}. Assuming
that flaton decay repopulates the neutralino density, we find that the
neutralino lifetime is $\approx 10^{6} - 10^{15}$ yr and falls in the
region forbidden by the astrophysical bounds \cite{ellis8}. We are
therefore forced to reject the flatino as a dark matter candidate, at
least in the form presented in ref.~\cite{ellis6}. There are few
phenomenological restrictions on cryptons at present since their
nominal cosmological abundance would be far in excess of cosmological
limits and must needs be substantially diluted by invoking entropy
generation subsequent to freeze-out. Hence their relic density is quite
uncertain \cite{ellis7}. Further, such particles can also decay through
non-renormalizable superpotential interactions and are thus constrained
by the aforementioned astrophysical bounds if their relic density
happens to be comparable to that of the dark matter \cite{ellis8}.
\smallskip \par
Inspired by the most general supergravity models \cite{lahanas}, we
shall break supersymmetry softly at a scale $m_{\X}$ with the initial
conditions of universal gaugino masses $m_{1/2}$, scalar masses $m_{0}$
and trilinear scalar couplings $A$. For the sake of expediency we
specifically assume $m_{\X} =  m_{\GUT}$. However, we argue that changing
the particular details of the model at high energies has only minor
consequences for our analysis, and that we are therefore addressing a
large class of possible theories. Consider, for example, the case where
$m_{\X} = 10^{-2} m_{\GUT}$. Between $m_{\X}$ and $m_{\GUT}$, the
gauginos and squarks receive contributions from the RGEs of $\sim 5\%$
of their final values, so that the assumed degeneracy of the soft
breaking parameters is approximately correct.  Likewise, we
expect the effect of the deconfinement of any hidden sector to be
mitigated if it happens reasonably close to the GUT scale.  In
addition, we observe the following: any parameters which do not depend
on the values of the strong coupling or top quark Yukawa coupling,
converge to infra-red stable points at low energy and are therefore
relatively independent of the (dimensionless) details of the model at
high energies. On the other hand parameters such as $\lambda_{7}$ which
{\em do} depend on the strong coupling or top quark Yukawa coupling,
are dominated by their (diverging) values at low energies and
are thus also expected to be relatively stable to changes in the
details of the model at high energy.
\smallskip \par
We shall only consider interactions relevant for neutralino masses
below that of the $\W$ boson. (Above the $\W$ mass, the annihilation
cross-section receives important contributions from processes such as
$\chi \chi \rightarrow {\W}^{+} {\W}^{-}$ \cite{griest1}). Since the
infra-red fixed point of $\lambda_{8}$ is only $\sim 0.21$
\cite{ellis3}, we expect our analysis to be valid for most of the
parameter space where $\langle \phi_{0} \rangle $ and $m_{1/2}$ are not
too large. When the neutralino mass exceeds $m_{\W}$, our results are
only qualitative.
\smallskip \par
If we neglect the bottom quark mass, then the squark and slepton mass matrices
which we need are already diagonal, and are as follows:
\begin{description}

\item the up and charm squarks,

\begin{equation}
\left(
\begin{array}{c}
 m^{2}_{\tilde{\up}_{\Lf}} \nonumber \\
 \nonumber \\
 m^{2}_{\tilde{\up}_{\R}}
\end{array}
\right)
 =
\left(
\begin{array}{c}
 m^{2}_{\Q} - g_{\Y}^{2} ({\bar v}^{2} - v^{2}) / 12 + g_{2}^{2}({\bar v}^{2} -
  v^{2}) / 4 \nonumber \\
 \nonumber \\
 m^{2}_{\Up} + g_{\Y}^{2} ({\bar v}^{2} - v^{2}) / 3
\end{array}
\right) ,
\end{equation}

\item the down, strange and bottom squarks,

\begin{equation}
\left(
\begin{array}{c}
 m^{2}_{\tilde{d}_{\Lf}} \nonumber \\
 \nonumber \\
 m^{2}_{\tilde{d}_{\R}}
\end{array}
\right)
 =
\left(
\begin{array}{c}
 m^{2}_{\Q} - g_{\Y}^{2} ({\bar v}^{2} - v^{2}) / 12 - g_{2}^{2}({\bar v}^{2} -
  v^{2}) / 4 \nonumber \\
 \nonumber \\
 m^{2}_{\Dn} - g_{\Y}^{2} ({\bar v}^{2} - v^{2}) / 6
\end{array}
\right) ,
\end{equation}

\item the left and right handed sleptons,

\begin{equation}
\left(
\begin{array}{c}
 m^{2}_{\tilde{\e}_{\Lf}} \nonumber \\
 \nonumber \\
 m^{2}_{\tilde{\e}_{\R}}
\end{array}
\right)
=
\left(
\begin{array}{c}
 m^{2}_{\Lf} + g_{\Y}^{2} ({\bar v}^{2} - v^{2}) / 4 - g_{2}^{2}({\bar v}^{2} -
  v^{2}) / 4 \nonumber \\
 \nonumber \\
  m^{2}_{\E} - g_{\Y}^{2} ({\bar v}^{2} - v^{2}) / 2
\end{array}\right) ,
\end{equation}

\item and, the sneutrino,

\begin{equation}
 m^{2}_{\tilde{\nu}_{\Lf}} = m^{2}_{\Lf} + g_{\Y}^{2} ({\bar v}^{2} - v^{2})/4
  + g_{2}^{2}({\bar v}^{2} - v^{2}) / 4\ ,
\end{equation}

\end{description}
\medskip
\noindent
where $v$, ${\bar v}$ and $x$ are the VEVs of the scalar fields
${\h}^0$, ${\bar {\h}}^0$ and $\phi_{0}$, and the rest of the notation
is as in ref.~\cite{lahanas}.
\smallskip \par
In addition we have three Higgs states, whose mass squared matrix is:

\begin{equation}
\left(
\begin{array}{lcr}
\! g^{2} v^{2}\! - \!\!(A_{7} \lambda_{7}\!+ \!3 \lambda_{7} \lambda_{8} x)
 {\bar v} x/v  &
 \!(2 \lambda_{7}^{2} \! -\!g^{2} )v {\bar v}\! + \! (A_{7}\!
 + \!3 \lambda_{8} x) \lambda_{7} x &
 \! (A_{7} \! + \! 6  \lambda_{8} x ) \lambda_{7}{\bar v}\!
 + \!  2 \lambda_{7}^{2} v x \! \nonumber \\
& & \nonumber \\

\! (2 \lambda_{7}^{2} \!  -\! g^{2} )v {\bar v}\! + \!(A_{7} \!
 + \!  3  \lambda_{8} x) \lambda_{7} x &
\! g^{2} {\bar v}^{2}\!  -\!  (A_{7} \lambda_{7}\!
 + \! 3 \lambda_{7} \lambda_{8} x) v x / {\bar v} &
 \! (A_{7} \! + \! 6  \lambda_{8} x)\lambda_{7} v \!+\! 2 \lambda_{7}^{2}
 {\bar v} x \!\nonumber \\
& &\nonumber \\

 \!  (A_{7} \! +\! 6  \lambda_{8} x ) \lambda_{7}{\bar v}\!
 + \! 2 \lambda_{7}^{2} v x  &
 \! (A_{7} \! + \!6  \lambda_{8} x)\lambda_{7} v\! +\! 2 \lambda_{7}^{2}
 {\bar v} x &
 \!3 A_{8} \lambda_{8} x\! + \!36 \lambda_{8}^{2} x^{2} \! -\!A_{7} \lambda_{7}
  v {\bar v}/x  \!
\end{array}
\right)
\end{equation}

\medskip
\noindent
where $g^{2} = (g^{2}_{\Y} + g_{2}^{2}) / 2$, and we define, for future
use: $\tan \beta_{x} \equiv x / \bar{v},\ \tan \beta \equiv {\bar{v}} /
v$. Finally there are two physical states coming from the pseudoscalar
mass squared matrix,

\begin{equation}
- \left(
\begin{array}{cc}
(A_{7}  + 3 \lambda_{8}  x) \lambda_{7} x V^{2}/v {\bar v}  &
  ( A_{7} \lambda_{7} -6 \lambda_{7} \lambda_{8} x) V \nonumber \\
& \nonumber \\
(A_{7} \lambda_{7} -6 \lambda_{7} \lambda_{8} x ) V  &
 ( A_{7}+ 12 \lambda_{8} x  )\lambda_{7} v {\bar v}/x + 9 A_{8} \lambda_{8} x
\end{array}
\right),
\end{equation}

\medskip
\noindent
where $V^{2}=(v^{2} + {\bar v}^{2})$. The $A_{7}$ and $A_{8}$ are the
coefficients of the $\lambda_{7}$ and $\lambda_{8}$ trilinear scalar
coupling terms. In order to be able to identify the Goldstone boson
state in the above, we have substituted in the minimisation conditions
\cite{ellis3}, and find the massless eigenvector to be

\begin{equation}
\tilde{G}=  \sin{\beta} {\bar {\h}} - \cos{\beta} \h\ ,
\end{equation}

\medskip
\noindent
as expected from consideration of current conservation. This use of the
minimisation condition has the added advantage that it enables one to
avoid estimating the SUSY breaking, gauge singlet mass-squared term,
which is the most sensitive to small changes in the Yukawa couplings.
As stated earlier, the remaining low energy parameters may be
approximated to better than $5\%$ using the RGEs as follows.

\smallskip \par
We begin at $m_{\W}$ and renormalize the Yukawa and gauge couplings to
$m_{\GUT}$ where $g_{3}$ and $g_{2}$ are unified. We have used 2-loop
RGEs for the gauge couplings, and 1-loop RGEs for all other parameters,
as in ref.~\cite{ellis5}. At each step in the renormalization
procedure, we keep track of the parameters by expanding them in terms
of the Yukawa couplings at $m_{\W}$. Then the supersymmetry breaking
terms $m_{1/2}$, $A$ and $m_{0}$ are switched on, and all the
parameters evolved back down to $m_{\W}$. This gives a set of
polynomials in $\lambda_{1}$, $\lambda_{2}$, $\lambda_{7}$,
$\lambda_{8}$, $m_{1/2}$, $m_{0}$ and $A$ which we truncate at sixth
order in the Yukawa couplings.
\smallskip \par
The current LEP values of the strong coupling (extracted from $\Z$ event
shapes using resummed QCD) \cite{bethke} and of the weak mixing angle
\cite{burkhardt} are given by :

\begin{equation}
 \alpha_{\s} (m_{\Z}) = 0.124 \pm 0.005\ , \qquad
 \sin^{2} \theta_{\W} (m_{\Z}) = 0.2325 \pm 0.0007\ ,
\end{equation}

\noindent
where both quantities are evaluated in the $\MS$ scheme, as is
appropriate for our renormalization group calculations. However,
analyses of deep inelastic $\mu,\,\nu$ scattering give a somewhat
smaller value:  $\alpha_{\s} (m_{\Z}) = 0.112 \pm 0.005$,
while analyses of $\tau$ decays and of J/$\Psi,\,\Upsilon$ decays give
intermediate values, all with comparable errors \cite{bethke}.
\noindent
Given the present uncertainty in determinations of $\alpha_{\s}$, we
prefer to perform our calculations using the `weighted mean' value
\cite{bethke}

\begin{equation}
 \alpha_{\s} (m_{\Z}) = 0.118 ,
\end{equation}

\noindent
with $\sin^{2} \theta_{\W} (m_{\Z}) = 0.2325$, and present the
corresponding results in Appendix A. (Note that the equivalent
expressions for the MSSM may be deduced by setting $\lambda_{7} =
\lambda_{8} = 0$, replacing $A_{7}$ by $B_\mu$, and the first term in
$B_\mu$ by $B m_{0}$, where $B$ is the quadratic scalar Higgs coupling
at the GUT scale.) In order to illustrate the effect of a small change
in $\alpha_{\s}$, we also solve the RGEs using $\alpha_{\s} = 1/8 =
0.125$ in Appendix B and using $\alpha_{\s} = 0.113$ in Appendix C. It
should be noted that since $\sqrt{\frac{5}{3}} g_{\Y}$ is still
significantly below $g_{2}$ at the GUT scale, the value of $m_{\Y}$ is
slight grey larger than suggested by eq.~(\ref{GUTvalue}). In fact the
correct relation is:

\begin{equation}
 m_{\Y} = \left( \frac{g_{2}^{2}(0)}{g_{\Y}^{2}(0)} \right)
  \left(\frac{g_{\Y}^{2}}{g_{2}^{2}} \right) m_{2}.
\end{equation}

\smallskip \par
These approximations turn out to be much better than expected, again
because of the convergence of the couplings to infra-red stable points.
Thus, despite the fact that the analytic approximations diverge from
the correct values during the running of the RGEs towards the GUT
scale, upon returning to the weak scale we find an accuracy of better
than $5\%$ when compared with numerical runnings, even close to the
unitarity limit. The results obtained are sensitive to the value of
$\alpha_{\s}$ and the current uncertainty ($\pm 0.005$) in its
determination corresponds to an uncertainty of $\sim 10\%$ in the
calculated couplings. We find a discrepancy of $\sim 10\%$ with the
results obtained using one-loop expressions for the gauge couplings
(which for the first two generations reproduce the numerical values
obtained in ref.~\cite{ellis2}). For the case $\alpha_{\s} = 1/8$, we
find good agreement with the numerical results of ref.~\cite{ibanez},
taking $\lambda_{2_{\tp}} = 0.31$.

\section{ Relic Abundance of the Neutralino }

The relic neutralino energy density may be computed by solution of the
Boltzmann transport equation governing the abundance of any stable
particle in the expanding universe. The exercise is greatly simplified
by noting that under suitable assumptions this reduces to the
continuity equation \cite{lee}

\begin{equation}
\label{continuity}
 \frac{{\dn}}{{\dn} t} (n_{\chi} R^3)= - {\langle \sigma_{\ann}\,v \rangle}
   \left[ n_{\chi}^{2} - (n_{\chi}^{\eq})^{2} \right] R^3\ ;
\end{equation}

\noindent
this simply states that the number of neutralinos (of density
$n_{\chi}$) in a comoving volume is governed by the competition between
annihilations and creations, the latter being dependent, by detailed
balance, on the equilibrium neutralino density, $n_{\chi}^{\eq}$. This
equation holds if the annihilating particles are non-relativistic and
if both the annihilating particles and the annihilation products are
maintained in kinetic equilibrium with the background thermal plasma
through rapid scattering processes \cite{bernstein,gondolo}. The
essential physical input is the `thermally-averaged' annihilation
cross-section, which may be conveniently calculated in the rest frame
of one of the annihilating particles (i.e. the `lab' frame)
\cite{gondolo}:

\begin{eqnarray}
\label{sigmav1}
 \langle \sigma_{\ann}\,v \rangle & = & \int_{0}^{\infty} {\dn} \epsilon\
  {\cal K} (x, \epsilon)\ \sigma_{\ann}\,v\ , \\ [1ex] \nonumber
 {\cal K} (x, \epsilon) & \equiv & \frac{2 x}{K_{2}^{2} (x)} \sqrt{\epsilon}\,
  (1 + 2 \epsilon)\ K_{1} (2 x \sqrt{1 + \epsilon})\ ,
 \quad x \equiv m_{\chi}/T\ ,
 \quad \epsilon = \frac{s - 4 m_{\chi}^{2}}{4 m_{\chi}^{2}}\ ,
\end{eqnarray}

\noindent
where $v$ is the relative velocity of the annihilating neutralinos, $s$
is the usual Mandelstam variable, and the $K_{n}$ are modified Bessel
functions of order $n$. This 1-dim integral can be easily evaluated
numerically and is well-behaved even if $\sigma_{\ann}\,v$ varies
rapidly with $\epsilon$, as near a resonance or at the opening up of a
new annihilation channel. (Useful analytic approximations have been
provided for these situations in refs.~\cite{gondolo,griest2}.) Outside
these problematic regions, a simple series expansion for $\langle
\sigma_{\ann}\,v \rangle$ is useful \cite{gondolo,srednicki}:

\begin{eqnarray}
  \langle \sigma_{\ann}\,v \rangle & = & a^{(0)} + \frac{3}{2} a^{(1)} x^{-1}
  + \left(\frac{9}{2} a^{(1)} + \frac{15}{8} a^{(2)} \right) x^{-2} \\
\nonumber
&&+ \left(\frac{15}{16} a^{(1)} + \frac{195}{16}a^{(2)} + \frac{35}{16}a^{(3)}
     \right) x^{-3} + \cdots
\end{eqnarray}

\noindent
where $a^{(n)}$ is the $n$th derivative of $\langle \sigma_{\ann}\,v
\rangle$ with respect to $\epsilon$, evaluated at $\epsilon = 0$. We will be
interested in the value of $\langle \sigma_{\ann}\,v \rangle$ for very
non-relativistic annihilating particles (with $x \sim 20$), hence it is
adequate to retain terms only upto ${\cal O}(x^{-1})$ if errors of a
few percent are tolerable. Therefore we write

\begin{equation}
\label{sigmav2}
\langle \sigma_{\ann}\,v \rangle \simeq a^{(0)} + \frac{3}{2} a^{(1)} x^{-1} .
\end{equation}

\noindent
Since the neutralino ($\chi$) is a Majorana fermion, $a^{(0)}$ (the
S-wave part in the above) is suppressed relative to $a^{(1)}$ (which
contains both S and P-wave pieces) due to the smallness of the final
state fermion masses \cite{goldberg}. As noted above, this expansion
breaks down close to resonances where the value of $a^{(1)}$ may even
be negative; however careful analysis \cite{gondolo,griest2} shows that
the true situation is not in fact pathological. We are mainly
interested in identifying the broad characteristics of parameter space,
eq.~(\ref{sigmav2}) is therefore adequate for our purposes. The above
expansion also breaks down close to particle thresholds because one of
the denominators in $a^{(1)}$ decreases to zero. In our case this
happens when the $\chi$ mass is close to that of the bottom quark;
however as we will see, such low masses for $m_{\chi}$ are in fact
disallowed by experiment. Finally, we stress again that only
annihilation channels relevant for $m_{\chi} \leqsim m_{\W}$ are
considered here. Above the $\W$ mass, $\langle \sigma_{\ann}\,v
\rangle$ receives important contributions from processes such as $\chi
\chi \rightarrow \W^{+} \W^{-}$, as discussed in ref.~\cite{griest1}.
This is relevant to the present discussion only for small values of
$\tan \beta_{x}$ and large values of $m_{1/2}$, where our results are
only qualitative.
\par
Most solutions to the continuity equation (\ref{continuity}) in the
literature (e.g. \cite{kolb,lee,bernstein}) have been obtained by
transforming variables from $t \rightarrow T$ assuming that the product
of the cosmic scale-factor $R$ and the photon temperature $T$ is
constant. However, $RT$ does change as the temperature in the cooling
universe falls below various mass thresholds and the corresponding
particles annihilate and release entropy (i.e. thermalized interacting
particles) into the system, according to

\begin{equation}
\label{RT}
 \frac{{\dn} R}{R} = -\frac{{\dn} T}{T} - \frac{1}{3}
    \frac{{\dn} g_{s_{\I}}}{g_{s_{\I}}} .
\end{equation}

\noindent
Here $g_{s_{\I}}$ counts the number of {\it interacting} degrees of freedom
which determine the specific entropy density, $s_{\I}$:

\begin{eqnarray}
\label{gs}
 g_{s_{\I}}(T) & \equiv & \frac{45}{2\pi^{2}} \frac{s_{\I}(T)}{T^3}\ , \\
 \nonumber \mbox{where,} \quad
 s_{\I} & = & \sum_{\rm int} g_{i} \int \frac{3 m_{i}^{2} + 4 p^{2}}
  {3 E_{i}(p)\,T}\,f_{i}^{\eq}(p, T)\,\frac{{\dn}^3 p}{(2\pi)^3}\ .
\end{eqnarray}

\noindent
Here the sum is over all interacting particles, $E_{i} \equiv
\sqrt{m_{i}^{2} + p^{2}}$, $g_{i}$ is the number of internal (spin)
degrees of freedom, and the equilibrium phase-space distribution
function (for zero chemical potential) is, $f_{i}^{\eq}(p, T) = [\exp
(E_{i}/T) \pm 1]^{-1}$, with $+/-$ corresponding to
Fermi/Bose statistics.
\smallskip \par
In addition we have to keep track of any particle species which
`decouples', i.e.  goes out of kinetic equilibrium with the thermal
plasma, while still relativistic, thus taking its entropy content out
of the `interacting' sector (which defines the photon temperature). The
quantity which remains truly invariant (in a comoving volume) is the
total entropy $S = s R^3$, where

\begin{equation}
\label{sdef}
 s (T) \equiv \frac{2\pi^{2}}{45} g_{s}(T)\,T^3\ ,
\end{equation}

\noindent
with $g_{s}$, the entropy degrees of freedom, related to $g_{s_{\I}}$ as
\cite{gondolo}

\begin{equation}
 g_{s} (T) = g_{s_{\I}} (T) \prod_{\rm dec}
               \left[1 + \frac{g_{s_j}(T_{{\D}_j})}{g_{s_{\I}}
                (T_{{\D}_j})}\right] \ ,
\end{equation}

\noindent
The product is over all the decoupled particle species $j$ (which
decouple at $T = T_{{\D}_j}$). Such particles will not share in any
subsequent entropy release hence their temperature will drop below that
of the interacting particles. By conservation of the total entropy, we
obtain \cite{gondolo,srednicki} :

\begin{equation}
 \frac{T_{j}}{T} = \left[\frac{g_{s_j} (T_{\D})}{g_{s_j}(T)}
  \frac{g_{s_{\I}} (T)}{g_{s_{\I}} (T_{\D})}\right]^{1/3} .
\end{equation}

\noindent
The total energy density includes, of course, all particles, decoupled
or interacting, appropriately weighted by their respective
temperatures. The energy degrees of freedom, $g_{\rho}$, is given by

\begin{eqnarray}
\label{grho}
 g_{\rho} (T) & \equiv & \frac{30}{\pi^{2}} \frac{\rho (T)}{T^{4}}\ , \\
  \nonumber
 \mbox{where,} \quad
 \rho & = & \sum_{\rm all} g_{k} \int E_{k}(q)\,f_{k}^{\eq}
  (p, T_{k})\,\frac{{\dn}^3 p}{(2\pi)^3}\ .
\end{eqnarray}

\noindent
In the present case the only particles which decouple while
relativistic are massless neutrinos, hence $g_{s}$ is actually the same as
$g_{s_{\I}}$ above the neutrino decoupling temperature of ${\cal
O}$~(MeV), and can replace the latter in eq.~(\ref{RT}) (see e.g.
ref.~\cite{srednicki})
\smallskip \par
The continuity equation (\ref{continuity}) can now be rewritten
\cite{gondolo} in terms of the dimensionless variables $Y_{\chi} \equiv
n_{\chi}/s$, $Y_{\chi}^{\eq} \equiv n^{\eq}_{\chi}/s$, and $x \equiv
m_{\chi}/T$ as

\begin{eqnarray}
\label{Yseqn}
 {{\dn} Y_{\chi} \over {\dn} x} & = & \lambda x^{-2} \left[
(Y_{\chi}^{\eq})^{2}
   - Y_{\chi}^{2} \right] , \\ \nonumber
 \mbox{where,} \quad
 \lambda & \equiv & \sqrt{\frac{\pi}{45}}\,m_{\chi}\,m_{\Pl}\,{\langle
  \sigma_{\ann}\,v \rangle}\,g_{\star}^{1/2}\ , \\ \nonumber \mbox{and,} \quad
 g_{\star}^{1/2}(T) & \equiv & \frac{g_{s}(T)}{\sqrt{g_{{\rho}}(T)}}
  \left(1 + \frac{1}{3} \frac{{\dn} \ln g_{s_{\I}}(T)}{{\dn} \ln T}
    \right)\ .
\end{eqnarray}

\noindent
The parameter $g_{\star}$ keeps track of the changing values of both
$g_{\rho}$ (which measures the total energy density and thus determines
the expansion rate $\dot{R}/R$), as well as $g_{s_{\I}}$ (which
determines the specific entropy density and consequently the value of
the adiabat $RT$).~\footnote{Warning: $g_{\star}$ is often used (e.g.
\cite{kolb}) to denote the {\em energy} degrees of freedom, which is
called $g_{\rho}$ here.} The values of $g_{\rho}(T)$ and
$g_{s_{\I}}(T)$ have been computed in ref.~\cite{srednicki} by explicit
integration over the particle distribution functions (using
eqs.~\ref{gs} and \ref{grho}), and $g_{\star}(T)$ has been calculated
from these data in ref.~\cite{gondolo}. There is considerable ambiguity
concerning the behaviour of these quantities during the quark-hadron
phase transition and the authors of ref.~\cite{srednicki} present
curves for two choices of the critical temperature, $T_{\crit}$, viz.
150 MeV and 400 MeV (corresponding to two choices of the `bag' constant
characterizing the confined hadrons). In fact the two curves merge at
$T \geqsim 800$ MeV, above which temperature it is appropriate to treat
the thermal plasma as an ideal gas of non-interacting fermions and
gauge bosons. As we shall see, the cosmologically interesting regions
of parameter space correspond to neutralino masses which are high
enough ($m_{\chi} \geqsim 20$ GeV) that freeze-out occurs well above
the quark-hadron phase transition; hence their relic abundance is not
particularly sensitive to the choice of $T_{\crit}$. To simplify the
presentation of our results, we specifically adopt the curves
corresponding to $T_{\crit} = 150$ MeV. Another source of ambiguity
concerns the electroweak phase transition, since if the relevant
critical temperature is small then this may alter the relic abundance
of massive particles which {\em obtain} their mass at this phase
transition \cite{dimopoulos}. The LEP bounds on the Higgs mass imply
that this does not happen in the minimal Standard Model
\cite{dimopoulos}, however in supersymmetric models, the electroweak
phase transition may well be strongly supercooled. Nonetheless this
ought not to affect the relic abundance of neutralinos lighter than a
few hundred GeV, which is of interest in this work. (However, such
effects ought to be taken into account for the heavier neutralinos,
$m_{\chi} \sim {\cal O}$ (TeV), considered in
refs.~\cite{olive1,griest1,olive2}.)
\smallskip \par
According to the continuity equation (\ref{Yseqn}), the particle
abundance tracks its equilibrium value as long as the self-annihilation
rate is sufficiently rapid relative to the expansion rate. When the
particle becomes non-relativistic, its equilibrium abundance falls
exponentially due to a Boltzmann factor, hence so does the annihilation
rate. Eventually, the annihilation rate becomes sufficiently small that
the particle abundance can no longer track its equilibrium value but
becomes constant (in a comoving volume). Hence the parameter

\begin{equation}
 \Delta \equiv {(Y_{\chi}- Y_{\chi}^{\eq}) \over Y_{\chi}^{\eq}}\ ,
\end{equation}

\noindent
grows exponentially from zero as the particle goes out of chemical equilibrium,
according to \cite{gondolo}

\begin{equation}
 \frac{45 g}{4 \pi^4} \frac{K_{2}(x)}{g_{s}} \lambda\ \Delta
  (\Delta + 2) = \frac{K_{2}(x)}{K_{1}(x)} - \frac{1}{x}
   \frac{{\dn} \ln g_{s_{\I}}}{{\dn} \ln T} ,
\end{equation}

\noindent
where we have used $Y_{\chi}^{\eq} = (45 g / 4\pi^4)[x^2 K_{2}(x) / g_{s}
(T)]$. This equation can be solved for the `freeze-out' temperature, $x
= x_{\fr}$ corresponding to a specific choice of $\Delta =
\Delta_{\fr}$; it is found that the choice $\Delta_{\fr} = 1.5$ gives a
good match to numerical solutions \cite{gondolo}. In fact, the
first term on the r.h.s. is $\sim 1$ for non-relativistic particles,
while the second term on the r.h.s. is zero and $g_{\star}^{1/2} =
g_{\rho}^{-1/2}$ if $g_{s_{\I}}$ is constant at the epoch of
freeze-out; the freeze-out temperature is then given by the more
commonly used condition (e.g. \cite{kolb}) :

\begin{eqnarray}
 x_{\fr} & = & \ln \left[\Delta_{\fr}(2 + \Delta_{\fr}) \delta \right]
  - {1 \over 2} \ln x_{\fr} \\ \nonumber
 \mbox{where} \quad \delta & = & \left(\frac{45}{32\pi^6}\right)^{1/2}
  g\,m_{\chi}\, m_{\Pl} {\langle \sigma_{\ann}\,v \rangle}\,g_{\rho}^{-1/2}\ .
\end{eqnarray}

\noindent
This corresponds to the epoch when the annihilation rate,
$n_{\chi}^{\eq} {\langle \sigma_{\ann}\,v \rangle}$, equals the rate of
change of the particle abundance itself \cite{lee}: ${\dn} \ln n^{\eq}
/ {\dn} t = x_{\fr}\ \dot{R} / R$. We find that for neutralinos,
$x_{\fr} \sim 20-25$.
\smallskip \par
After freeze-out, only annihilations are important since the
temperature is now too low for the inverse creations to proceed; hence
the asymptotic abundance, $Y_{\chi_\infty} \equiv Y_{\chi} (t \rightarrow
\infty)$, obtains by integrating ${{\dn} Y_{\chi} / {\dn} x} = -\lambda
x^{-2} Y_{\chi}^{2}$ with the initial condition, $Y_{\chi} (x_{\fr}) =
Y_{\chi}^{\eq} (x_{\fr})$ \cite{lee}. This gives

\begin{eqnarray}
 \frac{1}{Y_{\chi_\infty}} - \frac{1}{Y_{\chi} (x_{\fr})}
  & = & \int_{x_{\fr}}^{\infty} \frac{\lambda}{x^2} {\dn} x \\ \nonumber
  & = & \sqrt{\frac{45}{\pi}} m_{\Pl} \int_{0}^{T_{\fr}}
         \langle\sigma_{\ann}\,v \rangle\,g_{\ast}^{1/2}\ {\dn} T\ .
\end{eqnarray}

\noindent
Since $Y_{\chi_\infty}$ subsequently remains invariant, the number
density of the neutralinos today (at $T = T_{0}$) is obtained by simply
multiplying by the present entropy density, $s (T_{0})$, which is given
by eq.(\ref{sdef}) with $g_{s} (T_{0}) = 3.91$ and $T_{0} = 2.735 \pm
0.02\ ^{0}$~K \cite{mather}. The present neutralino
energy density is thus given by

\begin{equation}
  \Omega_{\chi}\,h^{2} = 2.775 \times 10^8\ Y_{\chi_\infty}
   \left(\frac{m_{\chi}}{\GeV}\right) .
\end{equation}

\noindent
The observational error in $T_{0}$ corresponds to an uncertainty of
$\pm 2 \%$ in $\Omega_{\chi}\,h^{2}$. The above procedure give an
accuracy of better than $5 \%$ when compared with numerical solutions
of eq.(\ref{Yseqn}) (e.g. \cite{srednicki}).
\smallskip \par
Two further remarks are in order. We have assumed above that the
neutralino, as the LSP, is absolutely stable by virtue of R-parity
invariance. However, most models have been considered as effective
theories derived from broken GUTs; in some cases (e.g the minimal
flipped SU(5) model) a small amount of R-parity breaking {\em is}
possible, causing the LSP to decay via superheavy boson exchange. As
noted earlier, if such metastable LSPs constitute the dark matter, then
constraints from astrophysical data \cite{ellis8} require their
lifetime to be at least $10^{6}$ times longer than the age of the
universe, hence they may be considered effectively stable.
\smallskip \par
We have also assumed that the neutralino is much lighter than the
charginos and sleptons, so that it cannot annihilate into these
particles at the epoch of freeze-out; such `co-annihilation' effects
can greatly reduce the relic abundance \cite{griest2}. In fact such
effects are important and must be taken into account for pure Higgsino
states ref.~\cite{mizuta}. However in the present case the neutralino
is either a gaugino or a mixed state over most of parameter space.
Also, the presence of the gauge singlet ensures that there is {\em no}
degeneracy between the neutralino and the squark masses as argued in
ref.~\cite{mizuta} for the case of the MSSM.

\section{ Exploring Parameter Space}

We now take the most general expression for the cross-section and
composition of the LSP and analyse the parameter space in the
$(\lambda_{7},\,\lambda_{8}),\ (m_{0},\,m_{1/2}),\ (\tan \beta,\, \tan
\beta_{x})$ and $(\tan \beta,\,A)$ planes.  Below the $\W$ mass, the
calculation of the cross-section is simplified since the contributing
diagrams are only those shown in fig.~1 (where $\f$ denotes a matter
fermion). The cross section may therefore be found as a general
expression involving left and right external fields and arbitrary
internal propagators; summing over all mass eigenstates and
interactions yields the required quantity (which is far too cumbersome
to be presented here). We have checked that our cross sections for
various pure LSP states agree with those found in
refs.~\cite{olive2,srednicki}.
\smallskip \par
We restrict our attention to SUSY breaking parameters less than $500$
GeV, to avoid excessive fine-tuning in the model.~\footnote{For a
model-independent survey see ref.~\cite{ross}.} As discussed earlier,
we look for cosmologically `interesting' regions in which the value of
the relic density is $0.1 \leqsim \Omega_{\chi}\,h^{2} \leqsim 0.5$,
and exclude regions in which $\Omega_{\chi}\,h^{2} \geqsim 1$. Using
recent experimental data from LEP \cite{davier} and CDF
\cite{laasanen}, we also exclude regions of parameter space where the
slepton, sneutrino, chargino, (lightest) Higgs or gluino have
masses which have already been excluded (or are less than that of the
neutralino), or where the supersymmetric contribution
{}~\cite{krauss,komatsu} to the (recently revised) ${\Z}^{0}$ `invisible
width' \cite{rolandi} is excessive. In particular, we impose the
following bounds (all at the $90 \%$ confidence level):

\begin{equation}
 m_{\tilde{l},\ \chi^\pm} > 45\ \GeV , \quad
 m_{\h} > 43\ \GeV , \quad
 m_{\tilde{\nu}} > 40\ \GeV , \quad
 m_{\tilde{g}} > 150\ \GeV, \quad
 (N_{\nu} - 3) < 0.11\ .
\end{equation}

\noindent
We also adopt $m_{\tp} = 120$ GeV as suggested by recent analyses of
high precision LEP data \cite{ellis9}. It is neccessary to exclude
negative Higgs and pseudo-Higgs masses-squared since otherwise the
absolute minimum of the potential would be CP-violating, as is implicit
in their mass matrices \cite{ellis3}.

\subsection{ The ($\lambda_{7}$, $\lambda_{8}$) plane }

Since the sfermion masses are independent of $\lambda_{7}$ and
$\lambda_{8}$ to first order (see the appendices), when we look in the
($\lambda_{7}$, $\lambda_{8}$) plane we see the effect on the LSP relic
density of the masses of the LSP, Higgs and pseudo-Higgs only. (In the
two remaining planes, however, the masses of all the (s)particles in
the low energy spectrum have an effect.) For a gaugino-like LSP,
sfermion exchange is the important contribution to $\langle \sigma_{\ann}\,v
\rangle$, but $\Z$ exchange becomes increasingly important for
Higgsino-like and mixed LSPs. Also, for Higgsino-like LSPs, the
sfermion exchange diagrams are suppressed by a factor of at least
$m_{\bt}/m_{\Z}$.
\smallskip \par
In contrast with previous work, it is not possible here to choose the
signs of $A_{7}$ and $A_{8}$, in order to guarantee the positiveness of the
Higgs and pseudo-Higgs masses-squared. Instead we choose the signs of
the VEVs such that the experimental conditions above are satisfied, and
allow $\lambda_{7}$ and $\lambda_{8}$ to be both positive and negative.
This translates to the following constraints on the VEVs:

\begin{eqnarray}
\label{vevs}
\mbox{Sgn}[{\bar{v}}]& = & +1 \nonumber \\
\mbox{Sgn}[v]        & = & - \mbox{Sgn}[\lambda_{7} \lambda_{8}] \nonumber \\
\mbox{Sgn}[x]        & = & + \mbox{Sgn}[A \lambda_{8}].
\end{eqnarray}

\noindent
We find no other choice of parameters which consistently meet all
experimental requirements. The first condition is chosen arbitrarily,
without loss of generality. Inspection of the neutral terms in the
potential shows that under the above conditions, all the masses are
invariant under the transformations $\lambda_{7} \rightarrow
-\lambda_{7}$ and $\lambda_{8} \rightarrow -\lambda_{8}$. We are
therefore free to choose $\lambda_{7}$ and $\lambda_{8}$ to be
positive.
\smallskip \par
In figs.~2a-d we show contours of the LSP mass and the relic density
in the ($\lambda_{7}-\lambda_{8}$) plane, as well as the experimentally
allowed regions. (Note that $\lambda_{7}$ and $\lambda_{8}$ are bounded
by their infra-red fixed values of $0.87$ and $0.21$ respectively). We
choose $m_{0}$ and $m_{1/2}$ to be comparable and of order the Fermi
scale since the changing LSP mass is then the dominant parameter in
determining the relic density. For small values of $\lambda_{7}$ and
$\lambda_{8}$, the LSP mass is small and the relic abundance large.
However these regions are unlikely to satisfy experimental bounds and
we find that this is indeed so. At large $\lambda_{7}$ and
$\lambda_{8}$, the Higgs and pseudo-Higgs masses increase and therefore
do not contribute appreciably to $\langle \sigma_{\ann}\,v \rangle$, which is
then dominated by sfermion exchange. Also the LSP becomes mostly
gaugino, and its mass becomes large and nearly constant. For positive
$A$, the relic density tends to a plateau (at a relatively low value)
in the experimentally allowed region (see fig.~2a). (The `holes' seen
in the figures correspond to poles in the annihilation cross-section
where the calculated relic density is neligibly small. As noted
earlier, this is an artifact of the approximate series expansion
eq.~(\ref{sigmav2}); correct thermal averaging of
the cross-section at resonances shows that the dips in the relic
abundance are not in fact as sharp or as deep \cite{gondolo,griest2}.)
\smallskip \par
The behaviour is similar, although somewhat more complex, for negative
$A$, but the more mixed LSP states produce a significantly lower relic
density in the experimentally allowed region. Higher values of $\Omega_{\chi}
h^{2}$ occur along the strips where $\lambda_{7} \sim 0.6$, and where
$\lambda_{7} \sim 0.25$, when the LSP mass is still relatively small
(see fig.~2b).  Because of the relations in eq.~(\ref{vevs}), the relic
density takes longer to reach its plateau at high values of the Yukawa
couplings since the LSP eigenstate is dominated by the mixed
higgsino/singlet.
\smallskip \par
In addition, making $|A|$, $\tan \beta$ or $m_{0}$ large causes the
relic density to flatten out more quickly, and tends to increase its
value in the flat region (see figs.~2c,d).  We find that the
experimentally allowed region tends to shrink for higher values of
$\tan \beta$ and $|A|$ and therefore conclude that these are
experimentally disfavoured.  These last two observations will be
discussed in more detail below.
\smallskip \par
For the values of parameters chosen in fig.~2d (i.e. $A = +1$, $\tan
\beta = 4$, $\tan \beta_{x} = 1/2$, $m_{\tp} = 120$ GeV, $m_{0} = 300$ GeV
and $m_{1/2} = 200$ GeV), $\Omega_{\chi}\,h^{2}$ flattens out at

\begin{equation}
 \Omega_{\chi}\,h^{2} \simeq 0.4\ ,
\end{equation}

\noindent
as seen in fig.~2e. In the experimentally allowed region, there is very
little variation with $\lambda_{8}$, for either positive or negative
$A$. In this region the neutralino mass lies in the range $\sim 60-80$ GeV.
\smallskip \par
Clearly the large value of $\lambda_{7}\ (= g_{2})$ often chosen in the
literature \cite{olive2} is in fact favoured by experiment; also the
relic density in this region is particularly constant. The value of
$\lambda_{8} \sim g_{2} / 6$ \cite{olive2} however seems to be too low,
and $\lambda_{8} \sim 0.2$ is a better choice. (In any case the former
value was derived using smaller values of the top quark coupling, and
we find that in our case the latter value is quite natural for
$\lambda_{2_{\tp}}$ of ${\cal O}(1)$ at the GUT scale.)
\smallskip \par
Henceforth, we shall choose $\lambda_{7} = g_{2}$ and $\lambda_{8} =
0.2$, to analyse the ($m_{0}$, $m_{1/2}$) and ($\tan \beta$, $\tan
\beta_{x}$) planes, bearing in mind that when the LSP contains a
significant proportion of higgsino or gauge singlet, the values
obtained for $\Omega_{\chi}\,h^{2}$may be slight grey reduced by choosing
larger $\lambda_{7}$.

\subsection{ The ($m_{0}$, $m_{1/2}$) plane }

In figs.~3a-d we exhibit contours of the relic density, together with
the experimentally allowed region, and the mass of the LSP (which
depends only on the value of $m_{1/2}$). There are three easily
identifiable regions in the ($m_{0}$, $m_{1/2}$) plane. The first is
the region close to the origin, where the relic density is small. This
is because here the sfermion, Higgs and pseudo-Higgs masses are small,
causing $\langle \sigma_{\ann}\,v \rangle$ to be correspondingly large.
The LSP mass increases linearly with $m_{1/2}$, until it becomes
dominated by the gauge singlet, when it becomes constant. At this point
the relic density tends to increase, as the sfermion exchange
contribution to $\langle \sigma_{\ann}\,v \rangle$ (the third diagram
of fig.~1) becomes suppressed. This increase becomes more marked for
larger $\tan \beta$, and as a result, the region of high
$m_{1/2}\ (\geqsim$ 300 GeV) appears to be excluded for these values.
The third region one can identify is where $m_{0}$ is large, the LSP
mass is small, and the Higgs and pseudo-Higgs masses are large. This
also yields a cosmologically interesting relic density for $m_{0}
\geqsim 200$ GeV, which is relatively independent of $\tan \beta$ and
$A$. This behaviour is similar to that found for the MSSM in
ref.~\cite{ellis2}. At low $m_{1/2}$ the LSP is mostly photino and has
a mass of order $\frac{2}{3} \sin^{2} \theta_{\W} m_{2} + {\cal O}
(m^{2}_{2} / m_{\Z})$, in accord with ref.~\cite{olive2}.
\smallskip \par
We find that high values of $\Omega_{\chi}\,h^{2}$ are favoured by high
values of $m_{0}$ and low values of $m_{1/2}$. In particular we note
that for $\tan \beta \geqsim4$ and negative $A$, there are no allowed
regions which are cosmologically interesting.  For positive $A$, the
no-scale theories ($m_{0} = 0$) are excluded by LEP as shown in
fig.~3a. Note that since we are using the choice of VEVs given by
eq.~(\ref{vevs}), the sign of $A$ may instead be regarded as a choice
of the sign of $\lambda_{8}$. For negative $A$ this is not the case,
and we find that there exists an upper limit on the allowed $m_{0}$,
coming from the requirement that the vacuum not be CP-violating (see
fig.~3b). This bound becomes more restrictive if the values of $\tan
\beta$ or $|A|$ are increased as seen in figs.~3c,d.

\subsection{ The ($\tan \beta,\ \tan \beta_{x}$) and ($\tan \beta,\ A$) planes
}

Choosing the above values of $\lambda_{7}$ and $\lambda_{8}$, we examine
the relic density in the ($\tan \beta$, $\tan \beta_{x}$) plane
(figs.~4a,b) for $A = \pm 1$, $m_{1/2} = 200$ GeV, and $m_{0} = 300$
GeV. The experimental constraints discussed earlier now place the
following limits on the allowed values of these parameters,

 \begin{equation}
\label{limits}
\begin{array}{rcl}
 0.5 \leqsim & \tan \beta \leqsim & 5.0 \nonumber \\
       & \tan \beta_{x} \leqsim & 2.0,
\end{array}
\end{equation}

\noindent
corresponding to the values of $\lambda_{7}$ and $\lambda_{8}$ chosen
here. (Higher values of $\tan \beta$ considerably shrink the
experimentally allowed regions in fig.~2.) These limits are
considerably less restrictive than those suggested in
ref.~\cite{olive2}, where the specific values chosen were $\lambda_{7}
= g_{2}$ and $\lambda_{8} = g_{2} / 6 \simeq 0.108$. In addition we
find that, remarkably enough, the parameter space is restricted to
those values of $\tan \beta$ and $\tan \beta_{x}$ which yield
$\Omega_{\chi} \sim 1$, and where the relic density is relatively
uniform, especially for positive $A$. In fig.~5, we examine the effect
of varying $A$ and $\tan \beta$. Now $\Omega_{\chi}\,h^{2}$ is
relatively constant over the whole plane, apart from at $A = 0$ where
we find a jump in the relic density from $\Omega_{\chi}\,h^{2} \sim
0.3$ down to $\Omega_{\chi}\,h^{2} \sim 0.1$ due to the change in the
sign of $\bar{v}$ and $v$ dictated by eq.~(\ref{vevs}). In general we
conclude that negative values of $A$ favour higher values of
$\Omega_{\chi}$. We can place additional limits on the trilinear
coupling,

\begin{equation}
\label{Abounds}
 -3 \leqsim A \leqsim 4 ,
\end{equation}

\noindent
which marginally favour positive values of $A$.

\section{ Conclusions }

Using a set of approximate analytic solutions to the RGEs, we have
examined the relic density of the LSP in a class of unified theories in
which the electroweak symmetry is broken by a gauge singlet. We have
used the most general expressions for all the sparticle masses and for
the thermally-averaged cross-section below the $\W$ mass. This is
particularly important when the LSP is gaugino-like, in which case the
dominant contribution to the cross-section is the sfermion exchange. In
addition, we stress the importance of using accurate expressions for
the scalar and trilinear coupling terms $A_{7}$ and $A_{8}$, since
these are involved in determining the electroweak breaking. The unknown
Yukawa couplings $\lambda_{7}$ and $\lambda_{8}$ are experimentally
restricted to have relatively high values,

\begin{equation}
\begin{array}{ccc}
 0.2 & \leqsim \lambda_{7} \leqsim & 0.8 \nonumber \\
 0.1 & \leqsim \lambda_{8} < & 0.21 ,
\end{array}
\end{equation}

\noindent
where the last limit is a unitarity constraint, above which the
na\"{\i}ve unification assumption is no longer valid. For the chosen
values of $m_{\tp} = 120$ GeV, $\lambda_{7} = g_{2}$, $\lambda_{8} =
0.2$, the values of the VEVs are constrained by eq.~(\ref{limits}) and
the trilinear coupling $A$ at the GUT scale by eq.~(\ref{Abounds}).
Further, positive values of $A$ are marginally favoured by experiment
and lead to a more uniform relic abundance in parameter space. We find
that cosmologically interesting regions exist with

\begin{equation}
 250 \mbox{ GeV} \leqsim m_{0} \leqsim 500 \mbox{ GeV} .
\end{equation}

\noindent
For negative $A$, higher values of $m_{0}$ tend to be excluded by
symmetry breaking requirements. The region $m_{1/2} \geqsim 300$ GeV is
excluded for larger $\tan \beta$, although one should now take into
account new processes involving $\W$ final states, and would therefore
expect $\Omega_{\chi}\,h^{2}$ to be smaller than our estimates in this
region.  As pointed out in ref.~\cite{olive2}, for various `pure' LSP
states (no gaugino content) there are cosmologically interesting
regions for much higher values of $m_{1/2}$, although such large masses
(exceeding 1 TeV) necessitate fine tuning of {\em at least} one part in
a hundred in the top quark Yukawa coupling, to obtain satisfactory
electroweak breaking \cite{ross}. (We also confirm the conclusion of
ref.~\cite{drees} that for a higgsino-like LSP (corresponding to large
$m_{1/2}$ and $m_{0}$ and small $\lambda_{8}$) there is {\it no} bound
on the LSP mass coming from the constraint $\Omega_{\chi}\,h^{2}
\leqsim 1$.) Although computational constraints dictated a tree-level
treatment of electroweak symmetry breaking in this work, we note that
1-loop corrections have a relatively minor effect on the relic density
\cite{olive2}. We therefore anticipate that our conclusions will remain
unaltered in a more detailed analysis of this model including radiative
corrections.

\vspace{1cm}
\noindent
{\bf\Large Acknowledgement} \hspace{0.3cm} We would like to thank Graham
Ross for useful discussions. SAA was supported by a SERC fellowship.

\newpage
\section*{Appendix A}

Analytic approximations for all required parameters (at $m_{\Z}$) in terms
of the Yukawa couplings (at $m_{\Z}$) and the supersymmetry-breaking terms
$m_{0}$, $m_{1/2}$ and $A$ for the minimally extended, supersymmetric,
standard model, with $\alpha_{\s} = 0.118$. Note that $m^{2}_{\Up}$,
$m^{2}_{\Q}$, $m^{2}_{\Dn}$, $\lambda_{1}$ and $\lambda_{2}$ carry suppressed
generation indices.~\footnote{The equivalent expressions for the MSSM
may be obtained by setting $\lambda_{7} = \lambda_{8} = 0$, replacing $A_{7}$
by $B_{\mu}$ and the first term in $B_{\mu}$ by $B m_{0}$, where $B$ is the
quadratic scalar Higgs coupling at the GUT scale.}

\begin{eqnarray*}
A_{1} & = & A m_{0} (1 \! - \! 0.78 \lambda_{1}^{2} \!
 - \! 0.13 \lambda_{2}^{2} \! - \! 0.31 \lambda_{7}^{2} \!
 + \! 0.03 \lambda_{1}^{2} \lambda_{7}^{2} \! - \! 0.02 \lambda_{2}^{2}
 \lambda_{7}^{2} \! + \! 0.25 \lambda_{7}^{2} \lambda_{8}^{2}) \\
 & + & m_{1/2}(- 3.85 \! + \! 1.69 \lambda_{1}^{2} \! + \! 0.01 \lambda_{1}^4
 \! + \! 0.28 \lambda_{2}^{2} \! + \! 0.11 \lambda_{7}^{2} \! + \! 0.04
 \lambda_{1}^{2} \lambda_{7}^{2} \! - \! 0.02 \lambda_{2}^{2} \lambda_{7}^{2}
 \! + \! 0.25 \lambda_{7}^{2} \lambda_{8}^{2}) \\ [1ex]
A_{2} & = & A m_{0} (1 \! - \! 0.13 \lambda_{1}^{2} \! - \! 0.77
 \lambda_{2}^{2} \! - \! 0.31 \lambda_{7}^{2} \!
 - \! 0.02 \lambda_{1}^{2} \lambda_{7}^{2} \! + \! 0.03 \lambda_{2}^{2}
 \lambda_{7}^{2} \! + \! 0.25  \lambda_{7}^{2} \lambda_{8}^{2} ) \\
 & + & m_{1/2} (-3.89 \! + \! 0.28 \lambda_{1}^{2} \! + \! 1.70 \lambda_{2}^{2}
 \! + \! 0.01 \lambda_{2}^4 \! + \! 0.11 \lambda_{7}^{2} \! - \! 0.02
 \lambda_{1}^{2} \lambda_{7}^{2}  \! + \! 0.04 \lambda_{2}^{2} \lambda_{7}^{2}
 \!  + \! 0.25 \lambda_{7}^{2} \lambda_{8}^{2}) \\ [1ex]
A_{7} & = & A m_{0}(1 \! - \! 0.39 \lambda_{1}^{2} \! - \! 0.39
 \lambda_{2}^{2} \! - \! 1.23 \lambda_{7}^{2} \! - \! 0.11 \lambda_{1}^{2}
 \lambda_{7}^{2} \! - \! 0.11 \lambda_{2}^{2} \lambda_{7}^{2} \! - \! 7.46
 \lambda_{8}^{2} \! - \! 0.52 \lambda_{7}^{2} \lambda_{8}^{2} \! ) \\
 & + & m_{1/2} (\! - \! 0.58 \! + \! 0.85  \lambda_{1}^{2} \! + \! 0.85
 \lambda_{2}^{2} \! + \! 0.43 \lambda_{7}^{2} \! - \! 0.10 \lambda_{1}^{2}
 \lambda_{7}^{2} \! - \! 0.10 \lambda_{2}^{2} \lambda_{7}^{2} \! - \! 0.50
 \lambda_{7}^{2} \lambda_{8}^{2}) \\ [1ex]
A_{8} & = & A m_{0}(1 \! - \! 1.84 \lambda_{7}^{2} \! - \! 0.22 \lambda_{1}^{2}
 \lambda_{7}^{2} \! - \! 0.22 \lambda_{2}^{2} \lambda_{7}^{2} \! - \! 22.4
 \lambda_{8}^{2} \! - \! 3.05 \lambda_{7}^{2} \lambda_{8}^{2} \! ) \\
 & + & m_{1/2}(0.69 \lambda_{7}^{2} \! - \! 0.21 \lambda_{1}^{2}
 \lambda_{7}^{2} \! - \! 0.21 \lambda_{2}^{2} \lambda_{7}^{2} \! - \! 3.01
 \lambda_{7}^{2} \lambda_{8}^{2} ) \\ [1ex]
m^{2}_{\Up} & = & m_{0}^{2} (1\! - \! 0.77 \lambda_{2}^{2} \! - \! 0.26 A^{2}
 \lambda_{2}^{2} \! + \! 0.03 A^{2} \lambda_{1}^{2} \lambda_{2}^{2} \!
 + \! 0.20 A^{2} \lambda_{2}^4 \! + \! 0.07 \lambda_{2}^{2} \lambda_{7}^{2}
 \! + \! 0.13 A^{2} \lambda_{2}^{2} \lambda_{7}^{2} ) \\
 & + & A m_{0} m_{1/2}(1.14 \lambda_{2}^{2} \! - \! 0.15 \lambda_{1}^{2}
 \lambda_{2}^{2} \! - \! 0.86 \lambda_{2}^4 \! - \! 0.25 \lambda_{2}^{2}
 \lambda_{7}^{2}) \\
 & + & m_{1/2}^{2} (6.27 \! - \! 3.21 \lambda_{2}^{2} \! + \!
 0.15 \lambda_{1}^{2} \lambda_{2}^{2} \! + \! 0.94 \lambda_{2}^4 \!
 - \! 0.07 \lambda_{2}^{2} \lambda_{7}^{2} ) \\ [1ex]
m^{2}_{\Q} & = & m_{0}^{2}(1 \! - \! 0.39 \lambda_{1}^{2} \! - \! 0.13 A^{2}
 \lambda_{1}^{2} \! + \! 0.10 A^{2} \lambda_{1}^4 \! - \! 0.39 \lambda_{2}^{2}
 \! - \! 0.13 A^{2} \lambda_{2}^{2} \! + \! 0.03 A^{2} \lambda_{1}^{2}
 \lambda_{2}^{2} \! + \! 0.10 A^{2} \lambda_{2}^4 \! + \! \\
 & & 0.03 \lambda_{1}^{2} \lambda_{7}^{2} \! + \! 0.06 A^{2} \lambda_{1}^{2}
 \lambda_{7}^{2} \! + \! 0.04 \lambda_{2}^{2} \lambda_{7}^{2} \! + \! 0.06
 A^{2} \lambda_{2}^{2} \lambda_{7}^{2} ) \\
 & + & A m_{0} m_{1/2}(0.56 \lambda_{1}^{2} \! - \!
 0.44 \lambda_{1}^4 \! + \! 0.57 \lambda_{2}^{2} \! - \!
 0.14 \lambda_{1}^{2} \lambda_{2}^{2} \! - \! 0.43 \lambda_{2}^4 \!
 - \! 0.13 \lambda_{1}^{2} \lambda_{7}^{2} \! - \! 0.13 \lambda_{2}^{2}
 \lambda_{7}^{2} ) \\
 & + & m_{1/2}^{2} ( 6.68 \! - \! 1.59 \lambda_{1}^{2} \!
 + \! 0.46 \lambda_{1}^4 \! - \! 1.60 \lambda_{2}^{2} \! + \! 0.16
 \lambda_{1}^{2} \lambda_{2}^{2} \! + \! 0.47 \lambda_{2}^4 \!
 - \! 0.03 \lambda_{1}^{2} \lambda_{7}^{2} \! - \! 0.03 \lambda_{2}^{2}
 \lambda_{7}^{2} ) \\ [1ex]
m^{2}_{\Dn} \! & = & \! m_{0}^{2} (1\! - \! 0.78 \lambda_{1}^{2} \!
 - \! 0.26 A^{2} \lambda_{1}^{2} \! + \!
 0.20 A^{2} \lambda_{1}^4 \! + \!
 0.03 A^{2} \lambda_{1}^{2} \lambda_{2}^{2} \! + \! 0.07 \lambda_{1}^{2}
 \lambda_{7}^{2} \! + \!
 0.13 A^{2}  \lambda_{1}^{2} \lambda_{7}^{2} ) \\
 & + & A m_{0} m_{1/2} (1.13 \lambda_{1}^{2} \! - \!
 0.87 \lambda_{1}^4 \! - \! 0.14 \lambda_{1}^{2} \lambda_{2}^{2}
 \! - \! 0.25 \lambda_{1}^{2} \lambda_{7}^{2}) \\
 & + & m_{1/2}^{2} (6.22 \! - \! 3.17  \lambda_{1}^{2} \! + \!
 0.93  \lambda_{1}^4 \! + \! 0.16 \lambda_{1}^{2} \lambda_{2}^{2} \!
 - \! 0.07 \lambda_{1}^{2} \lambda_{7}^{2} ) \\ [1ex]
m^{2}_{\E} & = & m_{0}^{2}\! + \! 0.15 m_{1/2}^{2} \\ [1ex]
m^{2}_{\Lf} & = & m_{0}^{2}\! + \! 0.51 m_{1/2}^{2} \\ [1ex]
m_{\Y} & = & 0.41 m_{1/2} \\ [1ex]
m_{2} & = & 0.79 m_{1/2} \\ [1ex]
m_{3} & = & 2.77 m_{1/2}
\end{eqnarray*}

\newpage
\section*{Appendix B}

As in Appendix A, with $\alpha_{\s} = 0.125$.

\begin{eqnarray*}
A_{1} & = & A m_{0} (1 \! - \! 0.76 \lambda_{1}^{2} \! - \! 0.12
 \lambda_{2}^{2} \! - \! 0.31 \lambda_{7}^{2} \! + \!
 0.04 \lambda_{1}^{2} \lambda_{7}^{2} \! - \! 0.02 \lambda_{2}^{2}
 \lambda_{7}^{2} \! + \! 0.26 \lambda_{7}^{2} \lambda_{8}^{2} ) \\
 & + & m_{1/2}(-4.09 \! + \! 1.75 \lambda_{1}^{2} \! + \! 0.01 \lambda_{1}^4
 \! + \! 0.29 \lambda_{2}^{2} \! + \! 0.11 \lambda_{7}^{2} \! + \! 0.04
 \lambda_{1}^{2} \lambda_{7}^{2} \! - \! 0.02 \lambda_{2}^{2} \lambda_{7}^{2}
 \!  + \! 0.26 \lambda_{7}^{2} \lambda_{8}^{2}) \\ [1ex]
A_{2} & = & A m_{0} (1 \! - \! 0.13  \lambda_{1}^{2} \!  - \! 0.75
 \lambda_{2}^{2} \! - \! 0.31 \lambda_{7}^{2} \! - \! 0.02  \lambda_{1}^{2}
 \lambda_{7}^{2} \! + \! 0.04 \lambda_{2}^{2} \lambda_{7}^{2} \! + \! 0.26
 \lambda_{7}^{2} \lambda_{8}^{2} ) \\
 & + & m_{1/2} (-4.13 \! + \! 0.29 \lambda_{1}^{2} \! + \! 1.75 \lambda_{2}^{2}
 \! + \! 0.01 \lambda_{2}^4 \! + \! 0.11 \lambda_{7}^{2} \! - \! 0.02
 \lambda_{1}^{2} \lambda_{7}^{2} \! + \! 0.04 \lambda_{2}^{2} \lambda_{7}^{2}
 \! + \! 0.26 \lambda_{7}^{2} \lambda_{8}^{2}) \\ [1ex]
A_{7} & = & A m_{0}(1 \! - \! 0.38 \lambda_{1}^{2} \! - \! 0.37
 \lambda_{2}^{2} \! - \! 1.23 \lambda_{7}^{2} \! - \! 0.11 \lambda_{1}^{2}
 \lambda_{7}^{2} \! - \! 0.11 \lambda_{2}^{2} \lambda_{7}^{2} \! - \! 7.51
 \lambda_{8}^{2} \! - \! 0.48 \lambda_{7}^{2} \lambda_{8}^{2}
 \! + \! 0.32   \lambda_{8}^4 ) \\
 & + & m_{1/2} (\! - \! 0.59 \! + \! 0.87  \lambda_{1}^{2}
 \! + \! 0.88  \lambda_{2}^{2} \! + \! 0.44 \lambda_{7}^{2} \! - \! 0.11
 \lambda_{1}^{2} \lambda_{7}^{2} \! - \! 0.11 \lambda_{2}^{2} \lambda_{7}^{2}
 \! - \! 0.52 \lambda_{7}^{2} \lambda_{8}^{2} ) \\ [1ex]
A_{8} & = & A m_{0}(1 \! - \! 1.85 \lambda_{7}^{2} \! - \! 0.22 \lambda_{1}^{2}
 \lambda_{7}^{2} \! - \! 0.22 \lambda_{2}^{2} \lambda_{7}^{2} \! - \! 22.5
 \lambda_{8}^{2} \! - \! 3.0 \lambda_{7}^{2} \lambda_{8}^{2} \! + \! 0.95
 \lambda_{8}^4) \\
 & + & m_{1/2}(0.66 \lambda_{7}^{2} \! - \! 0.22 \lambda_{1}^{2}
 \lambda_{7}^{2} \! - \! 0.22 \lambda_{2}^{2} \lambda_{7}^{2} \! - \!
 3.10 \lambda_{7}^{2} \lambda_{8}^{2} ) \\ [1ex]
m^{2}_{\Up} & = & m_{0}^{2} (1\! - \! 0.75 \lambda_{2}^{2} \! - \! 0.25 A^{2}
 \lambda_{2}^{2} \! + \! 0.03 A^{2} \lambda_{1}^{2} \lambda_{2}^{2} \!
 + \! 0.19 A^{2} \lambda_{2}^4 \! + \! 0.07 \lambda_{2}^{2} \lambda_{7}^{2}
 \! + \! 0.12 A^{2} \lambda_{2}^{2} \lambda_{7}^{2} ) \\
 & + & A m_{0} m_{1/2}(1.17 \lambda_{2}^{2} \! - \! 0.15 \lambda_{1}^{2}
 \lambda_{2}^{2} \! - \! 0.87 \lambda_{2}^4 \!
 - \! 0.26 \lambda_{2}^{2} \lambda_{7}^{2}) \\
 & + & m_{1/2}^{2} (6.90  \! - \! 3.44  \lambda_{2}^{2} \! + \!
 0.17  \lambda_{1}^{2}  \lambda_{2}^{2} \! + \! 1.01 \lambda_{2}^4 \!
 - \! 0.07 \lambda_{2}^{2} \lambda_{7}^{2} ) \\ [1ex]
m^{2}_{\Q} & = & m_{0}^{2}(1 \! - \! 0.38 \lambda_{1}^{2} \! - \! 0.13 A^{2}
 \lambda_{1}^{2} \! + \! 0.10 A^{2} \lambda_{1}^4  \! - \! 0.37
 \lambda_{2}^{2} \! - \! 0.12 A^{2} \lambda_{2}^{2} \! + \! 0.03 A^{2}
 \lambda_{1}^{2} \lambda_{2}^{2} \! + \! 0.10 A^{2} \lambda_{2}^4 \! + \! \\
 & & 0.04 \lambda_{1}^{2} \lambda_{7}^{2} \! + \! 0.06 A^{2} \lambda_{1}^{2}
 \lambda_{7}^{2} \! + \! 0.04 \lambda_{2}^{2} \lambda_{7}^{2} \! + \! 0.06
A^{2}
 \lambda_{2}^{2} \lambda_{7}^{2} ) \\
 & + & A m_{0} m_{1/2}(0.58 \lambda_{1}^{2} \! - \! 0.44 \lambda_{1}^4 \!
 + \! 0.58 \lambda_{2}^{2} \! - \! 0.15 \lambda_{1}^{2} \lambda_{2}^{2} \!
 - \! 0.43 \lambda_{2}^4 \! - \! 0.13 \lambda_{1}^{2} \lambda_{7}^{2} \!
 - \! 0.13 \lambda_{2}^{2} \lambda_{7}^{2} ) \\
 & + & m_{1/2}^{2} ( 7.30 \! - \! 1.71 \lambda_{1}^{2} \! + \! 0.50
 \lambda_{1}^4 \! - \! 1.72 \lambda_{2}^{2} \! + \! 0.17 \lambda_{1}^{2}
 \lambda_{2}^{2} \! + \! 0.50 \lambda_{2}^4 \! - \! 0.03 \lambda_{1}^{2}
 \lambda_{7}^{2} \! - \! 0.03 \lambda_{2}^{2} \lambda_{7}^{2} ) \\ [1ex]
m^{2}_{\Dn} \! & = & \! m_{0}^{2} (1\! - \! 0.76  \lambda_{1}^{2} \! - \! 0.25
 A^{2} \lambda_{1}^{2} \! + \! 0.19 A^{2} \lambda_{1}^4 \! + \! 0.03 A^{2}
 \lambda_{1}^{2} \lambda_{2}^{2} \! + \! 0.07 \lambda_{1}^{2} \lambda_{7}^{2}
 \! + \! 0.13 A^{2} \lambda_{1}^{2} \lambda_{7}^{2} ) \\
 & + & A m_{0}  m_{1/2} (1.16  \lambda_{1}^{2} \! - \! 0.88 \lambda_{1}^4
 \! - \! 0.14 \lambda_{1}^{2}  \lambda_{2}^{2} \! - \! 0.26 \lambda_{1}^{2}
 \lambda_{7}^{2} ) \\
 & + & m_{1/2}^{2} (6.84 \! - \! 3.41 \lambda_{1}^{2} \! + \!
 1.0 \lambda_{1}^4 \! + \! 0.17 \lambda_{1}^{2} \lambda_{2}^{2} \!
 - \! 0.07 \lambda_{1}^{2} \lambda_{7}^{2} ) \\ [1ex]
m^{2}_{\E} & = & m_{0}^{2}\! + \! 0.15 m_{1/2}^{2} \\ [1ex]
m^{2}_{\Lf} & = & m_{0}^{2}\! + \! 0.52 m_{1/2}^{2} \\ [1ex]
m_{\Y} & = & 0.40 m_{1/2} \\ [1ex]
m_{2} & = & 0.78 m_{1/2} \\ [1ex]
m_{3} & = & 2.89 m_{1/2}
\end{eqnarray*}

\newpage
\section*{Appendix C}

As in Appendix A, with $\alpha_{\s} = 0.113$.

\begin{eqnarray*}
A_{1} & = & A m_{0} (1 \! - \! 0.80 \lambda_{1}^{2} \! - \! 0.13
 \lambda_{2}^{2} \! - \! 0.30 \lambda_{7}^{2} \! + \!
 0.03 \lambda_{1}^{2} \lambda_{7}^{2} \! - \! 0.02 \lambda_{2}^{2}
 \lambda_{7}^{2} \! + \! 0.24 \lambda_{7}^{2} \lambda_{8}^{2} ) \\
 & + & m_{1/2}(-3.65 \! + \! 1.64 \lambda_{1}^{2} \! + \! 0.01 \lambda_{1}^4
 \! + \! 0.28 \lambda_{2}^{2} \! + \! 0.10 \lambda_{7}^{2} \! + \! 0.04
 \lambda_{1}^{2} \lambda_{7}^{2} \! - \! 0.02 \lambda_{2}^{2} \lambda_{7}^{2}
 \!  + \! 0.24 \lambda_{7}^{2} \lambda_{8}^{2}) \\ [1ex]
A_{2} & = & A m_{0} (1 \! - \! 0.13  \lambda_{1}^{2} \!  - \! 0.79
 \lambda_{2}^{2} \! - \! 0.30 \lambda_{7}^{2} \! - \! 0.02  \lambda_{1}^{2}
 \lambda_{7}^{2} \! + \! 0.03 \lambda_{2}^{2} \lambda_{7}^{2} \! + \! 0.24
 \lambda_{7}^{2} \lambda_{8}^{2} ) \\
 & + & m_{1/2} (-3.69 \! + \! 0.27 \lambda_{1}^{2} \! + \! 1.65 \lambda_{2}^{2}
 \! + \! 0.01 \lambda_{2}^4 \! + \! 0.10 \lambda_{7}^{2} \! - \! 0.02
 \lambda_{1}^{2} \lambda_{7}^{2} \! + \! 0.04 \lambda_{2}^{2} \lambda_{7}^{2}
 \! + \! 0.24 \lambda_{7}^{2} \lambda_{8}^{2}) \\ [1ex]
A_{7} & = & A m_{0}(1 \! - \! 0.40 \lambda_{1}^{2} \! - \! 0.39
 \lambda_{2}^{2} \! - \! 1.21 \lambda_{7}^{2} \! - \! 0.11 \lambda_{1}^{2}
 \lambda_{7}^{2} \! - \! 0.11 \lambda_{2}^{2} \lambda_{7}^{2} \! - \! 7.33
 \lambda_{8}^{2} \! - \! 0.47 \lambda_{7}^{2} \lambda_{8}^{2}
 \! + \! 0.15   \lambda_{8}^4 ) \\
 & + & m_{1/2} (\! - \! 0.57 \! + \! 0.82  \lambda_{1}^{2}
 \! + \! 0.83  \lambda_{2}^{2} \! + \! 0.42 \lambda_{7}^{2} \! - \! 0.10
 \lambda_{1}^{2} \lambda_{7}^{2} \! - \! 0.10 \lambda_{2}^{2} \lambda_{7}^{2}
 \! - \! 0.49 \lambda_{7}^{2} \lambda_{8}^{2} ) \\ [1ex]
A_{8} & = & A m_{0}(1 \! - \! 1.82 \lambda_{7}^{2} \! - \! 0.21 \lambda_{1}^{2}
 \lambda_{7}^{2} \! - \! 0.21 \lambda_{2}^{2} \lambda_{7}^{2} \! - \! 22.0
 \lambda_{8}^{2} \! - \! 2.84 \lambda_{7}^{2} \lambda_{8}^{2} \! + \! 0.46
 \lambda_{8}^4) \\
 & + & m_{1/2}(0.62 \lambda_{7}^{2} \! - \! 0.21 \lambda_{1}^{2}
 \lambda_{7}^{2} \! - \! 0.21 \lambda_{2}^{2} \lambda_{7}^{2} \! - \!
 2.87 \lambda_{7}^{2} \lambda_{8}^{2} ) \\ [1ex]
m^{2}_{\Up} & = & m_{0}^{2} (1\! - \! 0.79 \lambda_{2}^{2} \! - \! 0.26 A^{2}
 \lambda_{2}^{2} \! + \! 0.04 A^{2} \lambda_{1}^{2} \lambda_{2}^{2} \!
 + \! 0.20 A^{2} \lambda_{2}^4 \! + \! 0.07 \lambda_{2}^{2} \lambda_{7}^{2}
 \! + \! 0.13 A^{2} \lambda_{2}^{2} \lambda_{7}^{2} ) \\
 & + & A m_{0} m_{1/2}(1.10 \lambda_{2}^{2} \! - \! 0.14 \lambda_{1}^{2}
 \lambda_{2}^{2} \! - \! 0.86 \lambda_{2}^4 \!
 - \! 0.24 \lambda_{2}^{2} \lambda_{7}^{2}) \\
 & + & m_{1/2}^{2} (5.74   \! - \! 2.99  \lambda_{2}^{2} \! + \!
 0.15  \lambda_{1}^{2}  \lambda_{2}^{2}   \! + \! 0.89  \lambda_{2}^4 \!
 - \! 0.06 \lambda_{2}^{2} \lambda_{7}^{2} ) \\ [1ex]
m^{2}_{\Q} & = & m_{0}^{2}(1 \! - \! 0.40 \lambda_{1}^{2} \! - \! 0.13 A^{2}
 \lambda_{1}^{2} \! + \! 0.11 A^{2} \lambda_{1}^4  \! - \! 0.39
 \lambda_{2}^{2} \! - \! 0.13 A^{2} \lambda_{2}^{2} \! + \! 0.03 A^{2}
 \lambda_{1}^{2} \lambda_{2}^{2} \! + \! 0.10 A^{2} \lambda_{2}^4 \! + \! \\
 & & 0.03 \lambda_{1}^{2} \lambda_{7}^{2} \! + \! 0.06 A^{2} \lambda_{1}^{2}
 \lambda_{7}^{2} \! + \! 0.03 \lambda_{2}^{2} \lambda_{7}^{2} \! + \! 0.06
A^{2}
 \lambda_{2}^{2} \lambda_{7}^{2}  ) \\
 & + & A m_{0} m_{1/2}(0.55 \lambda_{1}^{2} \! - \! 0.43 \lambda_{1}^4 \!
 + \! 0.55 \lambda_{2}^{2} \! - \! 0.14 \lambda_{1}^{2} \lambda_{2}^{2} \!
 - \! 0.43 \lambda_{2}^4 \! - \! 0.12 \lambda_{1}^{2} \lambda_{7}^{2} \!
 - \! 0.12 \lambda_{2}^{2} \lambda_{7}^{2} ) \\
 & + & m_{1/2}^{2} ( 6.15 \! - \! 1.48 \lambda_{1}^{2} \! + \! 0.44
 \lambda_{1}^4 \! - \! 1.50 \lambda_{2}^{2} \! + \! 0.15 \lambda_{1}^{2}
 \lambda_{2}^{2} \! + \! 0.44 \lambda_{2}^4 \! - \! 0.03 \lambda_{1}^{2}
 \lambda_{7}^{2} \! - \! 0.03 \lambda_{2}^{2} \lambda_{7}^{2} ) \\ [1ex]
m^{2}_{\Dn} \! & = & \! m_{0}^{2} (1\! - \! 0.80  \lambda_{1}^{2} \! - \! 0.27
 A^{2} \lambda_{1}^{2} \! + \! 0.21 A^{2} \lambda_{1}^4 \! + \! 0.03 A^{2}
 \lambda_{1}^{2} \lambda_{2}^{2} \! + \! 0.07 \lambda_{1}^{2} \lambda_{7}^{2}
 \! + \! 0.13 A^{2} \lambda_{1}^{2} \lambda_{7}^{2} ) \\
 & + & A m_{0}  m_{1/2} (1.09  \lambda_{1}^{2} \! - \! 0.86 \lambda_{1}^4
 \! - \! 0.14 \lambda_{1}^{2}  \lambda_{2}^{2} \! - \! 0.24 \lambda_{1}^{2}
 \lambda_{7}^{2} ) \\
 & + & m_{1/2}^{2} (5.70 \! - \! 2.97 \lambda_{1}^{2} \! + \!
 0.88 \lambda_{1}^4 \! + \! 0.15 \lambda_{1}^{2} \lambda_{2}^{2} \!
 - \! 0.06 \lambda_{1}^{2} \lambda_{7}^{2} ) \\ [1ex]
m^{2}_{\E} & = & m_{0}^{2}\! + \! 0.15 m_{1/2}^{2} \\ [1ex]
m^{2}_{\Lf} & = & m_{0}^{2}\! + \! 0.50 m_{1/2}^{2} \\ [1ex]
m_{\Y} & = & 0.42 m_{1/2} \\ [1ex]
m_{2} & = & 0.79 m_{1/2} \\ [1ex]
m_{3} & = & 2.67 m_{1/2}
\end{eqnarray*}

\newpage
\section*{Figures}

\begin{description}

\item{\bf Figure 1 } Diagrams contributing to the thermally-averaged
cross-section.

\item{\bf Figure 2a } Contour plots of $\Omega_{\chi}\,h^{2}$ and the
LSP mass (in GeV) in the $\lambda_{7} - \lambda_{8}$ plane, for $A =
+1$, $\tan \beta = 2$, $\tan \beta_{x} = 1/2$, $m_{\tp} = 120$ GeV,
$m_{0} = 300$ GeV and $m_{1/2} = 200$ GeV. The cosmologically
interesting region $0.1 < \Omega_{\chi}\,h^{2} < 0.5$ is shaded light
grey while the forbidden region $\Omega_{\chi}\,h^{2} > 1$ is shaded
black; the dark grey regions correspond to intermediate values of
$\Omega_{\chi}\,h^{2}$. The bold hatched line encloses the
experimentally allowed region. The arrows on the axes indicate the
infra-red fixed points.

\item{\bf Figure 2b } As in fig.~2a, but with $A = -1$.

\item{\bf Figure 2c } As in fig.~2a, but with $A = -3$.

\item{\bf Figure 2d } As in fig.~2a, but with $\tan \beta = 4$.

\item{\bf Figure 2e } $\Omega_{\chi}\,h^{2}$ versus $\lambda_{7}$ with
$\lambda_{8} = 0.2$; the other parameters are as in fig.~2d ($A = +1$,
$\tan \beta = 4$, $\tan \beta_{x} = 1/2$, $m_{\tp} = 120$ GeV, $m_{0} =
300$ GeV and $m_{1/2} = 200$ GeV). The experimentally allowed region is
demarcated by the vertical hatched lines.

\item{\bf Figure 3a } Contour plots of $\Omega_{\chi}\,h^{2}$ and the
LSP mass (in GeV) in the $m_{0} - m_{1/2}$ plane, for $A = +1$, $\tan
\beta = 2$, $\tan \beta_{x} = 1/2$, $\lambda_{7} = g_{2}$, $\lambda_{8}
= 0.2$ and $m_{\tp} = 120$ GeV. The cosmologically interesting region
$0.1 < \Omega_{\chi}\,h^{2} < 0.5$ is shaded light grey while the
forbidden region $\Omega_{\chi}\,h^{2} > 1$ is shaded black; the dark
grey regions correspond to intermediate values of
$\Omega_{\chi}\,h^{2}$.  The bold hatched line encloses the
experimentally allowed region.

\item{\bf Figure 3b } As in fig.~3a, but with $A = -1$.

\item{\bf Figure 3c } As in fig.~3a, but with $A = -3$.

\item{\bf Figure 3d } As in fig.~3b, but with $\tan \beta = 4$.

\item{\bf Figure 3e } $\Omega_{\chi}\,h^{2}$ versus $m_{0}$ with
$m_{1/2} = 200$ GeV; the other parameters are as in fig.~3b ($A = -1$,
$\tan \beta = 2$, $\tan \beta_{x} = 1/2$, $\lambda_{7} = g_{2}$,
$\lambda_{8} = 0.2$ and $m_{\tp} = 120$ GeV). The experimentally
allowed region is demarcated by the vertical hatched lines.

\item{\bf Figure 4a } Contour plots of $\Omega_{\chi}\,h^{2}$ and the
LSP mass (in GeV) in the $\tan \beta - \tan \beta_{x}$ plane, for $A =
+1$, $\lambda_{7} = g_{2}$, $\lambda_{8} = 0.2$, $m_{\tp} = 120$ GeV,
$m_{0} = 300$ GeV and $m_{1/2} = 200$ GeV. The cosmologically
interesting region $0.1 < \Omega_{\chi}\,h^{2} < 0.5$ is shaded light
grey while the forbidden region $\Omega_{\chi}\,h^{2} > 1$ is shaded
black; the dark grey regions correspond to intermediate values of
$\Omega_{\chi}\,h^{2}$.  The bold hatched line encloses the
experimentally allowed region.

\item{\bf Figure 4b } As in fig.~4a, but with $A = -1$.

\item{\bf Figure 5 } Contour plots of $\Omega_{\chi}\,h^{2}$ and the
LSP mass (in GeV) in the $\tan \beta - A$ plane, for $\tan \beta_{x} =
1/2$, $\lambda_{7} = g_{2}$, $\lambda_{8} = 0.2$, $m_{\tp} = 120$ GeV,
$m_{0} = 300$ GeV and $m_{1/2} = 200$ GeV. The cosmologically
interesting region $0.1 < \Omega_{\chi}\,h^{2} < 0.5$ is shaded light grey
while the forbidden region $\Omega_{\chi}\,h^{2} > 1$ is shaded black;
the dark regions correspond to intermediate values of $\Omega_{\chi}\,h^{2}$.
The bold hatched line encloses the experimentally allowed region.

\end{description}

\end{document}